\documentclass{ws-ijmpa}

\begin{document}

\markboth{D.Sokhan -- CLAS Collaboration}
{Beam Asymmetry in Pion Photoproduction from Neutron}

%
\catchline{}{}{}{}{}
%

\title{A NEW MEASUREMENT OF BEAM ASYMMETRY IN PION PHOTOPRODUCTION FROM THE NEUTRON USING CLAS}

\author{\footnotesize D.Sokhan$^{\star,}$\footnote{E-mail address: 
daria@jlab.org}~,
D.Watts$^{\star}$,
D.Branford$^{\star}$,
F.Klein$^{\%}$
and the CLAS Collaboration. 
}

\address{
$^{\star}$School of Physics \& Astronomy, University of Edinburgh, Edinburgh, UK\\ 
$^{\%}$Department of Physics, CUA, Washington, DC, USA\\
}

\maketitle


\begin{abstract}

We present a preliminary analysis of the photon beam asymmetry observable ($\Sigma$) from the photoproduction reaction channel $\gamma n \rightarrow \pi^{-} p$. This new data was obtained using the near-$4\pi$ CEBAF Large Acceptance Spectrometer (CLAS) at Jefferson Laboratory, USA, employing a linearly polarised photon beam with an energy range 1.1 - 2.3 GeV. The measurement will provide new data to address the poorly established neutron excitation spectrum and will greatly expand the sparse world data-set both in energy and angle.

\keywords{photonuclear; neutron; beam asymmetry.}
\end{abstract}

\section{Search for Nucleon Resonances}  

Despite decades of study our knowledge of the fundamental resonance spectrum of the nucleon is still incomplete. Many resonance properties are uncertain and some resonances predicted to exist are yet to be observed. Most predictions of the resonance spectrum are based on phenomenological models, such as the constituent quark model or the di-quark model, but lattice predictions directly from Quantum Chromodynamics (QCD) are fast developing. Experimental differentiation between these theoretical approaches based on the presently established resonance spectrum is, however, inconclusive due to the many ``missing'' and poorly established resonances.  Insufficient accuracy and quantity of observables obtained results in too many ambiguities and the famous case of the missing resonances in baryon spectroscopy \cite{Burkert}. Have they not yet been observed -- or are they simply not there?

\subsection{Pion Photoproduction and Beam Asymmetry}

A promising tool to learn about the resonance spectrum is the photoproduction of pions from nucleon targets, as many resonances are expected to couple to the pion decay channel and polarised real photons, with a well-understood electromagnetic interaction, provide a powerful probe.

In such pseudo-scalar meson production reactions the measurement of differential cross section, all three single-polarisation (beam, target, recoil nucleon) and a total of four well chosen double-polarisation observables has the potential to give the first model independent extraction of the production amplitudes. This will greatly enhance our capabilities to determine the nucleon excitation spectrum in a nearly model-independent way. \cite{BDS}$^,$\cite{Tabakin}. This effort is being carried out at various electron-beam facilities worldwide. 

One of the three single-polarisation observables, and a crucial one to constrain PWA's, is the Beam Asymmetry, $\Sigma$, from linearly polarised photons: 
\begin{equation}
\frac{d\sigma}{d\Omega} = \sigma_0(1-P_{lin} \Sigma \cos2\phi)
\label{asym_eq}
\end{equation}
where $P_{lin}$ is the linear beam polarisation and $\phi$ is the angle between the reaction plane and the photon polarisation plane. Data from the neutron is crucial in determining the resonance isospin and achieving a reliable extraction of the electromagnetic couplings of the excited states. The world dataset on the neutron, however, is extremely sparse, consisting of only three fixed-angle experiments limited to the 1.70 - 2.05 GeV range in energy and 35$^{\circ}$ - 90$^{\circ}$ in scattering angle. \cite{exp1}$^,$\cite{exp2}$^,$\cite{exp3}. 

\subsection{The g13 experiment}

A new, extensive photoproduction experiment using a polarised photon beam and a liquid deuterium target has been carried out at Jefferson Laboratory, Virginia, USA, March - June 2007. The experiment ran with both a circularly and a linearly polarised beam. The linear polarisation plane was flipped frequently between two orthogonal orientations, parallel and perpendicular to the lab floor. Six photon energy settings were employed between 1.1 and 2.3 GeV (produced via coherent bremsstrahlung from an electron beam at 3.3 - 5.2 GeV). A single charged particle trigger was used, yielding a total of $3 \times 10^{10}$ events recorded. We present very preliminary measurements of $\Sigma$ from the analysis of the $\gamma + n \rightarrow \pi^- + p$ channel.

\section{Experimental Facility -- Jefferson Lab}

Jefferson Laboratory is home to a 1.4 km race-track electron accelerator operating at energies up to 6 GeV (an upgrade to 12 GeV is currently underway). 200 $\mu$A of continuous current is split and delivered simultaneously to three experimental halls.
Our experiment was conducted in Hall B, where the electron beam is passed through a diamond crystal emitting linearly polarised photons via the process of coherent bremsstrahlung.

The scattered electrons are momentum analysed in the Hall B photon tagging facility\cite{tagger}, which serves to "tag" the photons through a coincidence timing measurement between the deflected electrons and the event time in the detector. The target cell is positioned in the centre of CLAS.

CLAS \cite{CLAS} is a formidable multi-layer onion of detectors (scintillators, drift chambers and calorimeters) providing nearly full coverage in the azimuthal angle and from 8$^\circ$ to 140$^\circ$ in scattering angle (lab frame) which, combined with a toroidal magnetic field, offer excellent sensitivity to charged particles.

\section{Analysis Techniques}

\subsection{Identifying the $\gamma + n \rightarrow \pi^- + p$ channel}
 
The reaction of interest can be considered as quasi-free with a spectator proton:
\begin{equation}
\gamma + d \rightarrow \pi^- + p + (p_{spectator})
\label{reaction_eq}
\end{equation}
The following data cuts are then applied to obtain a first selection of the $\gamma + n \rightarrow \pi^- + p$ channel:
\begin{itemlist}
  \item Select two-particle events with particle masses of the $\pi^-$ and {\it p} in the final state.
  \item Cut on ``missing mass'' of the recoiling system to be consistent with the mass of a spectator proton.
  \item Select low ``missing momentum'' in the spectator system - below 0.15 GeV/c.
  \item Cut on {\it p} and $\pi^-$ being back-to-back in centre-of-mass system (CMS).
\end{itemlist}

The exact photon was identified through timing coincidence of the reaction in CLAS with the tagger. This is made possible by the bunched nature of the beam, which arrives at 2 ns intervals.

\subsection{Extracting the Beam Asymmetry}

\begin{figure}[t]
\begin{center}
\begin{minipage}[t]{0.48\textwidth}
\centerline{\epsfysize 4.7 cm
\epsfbox{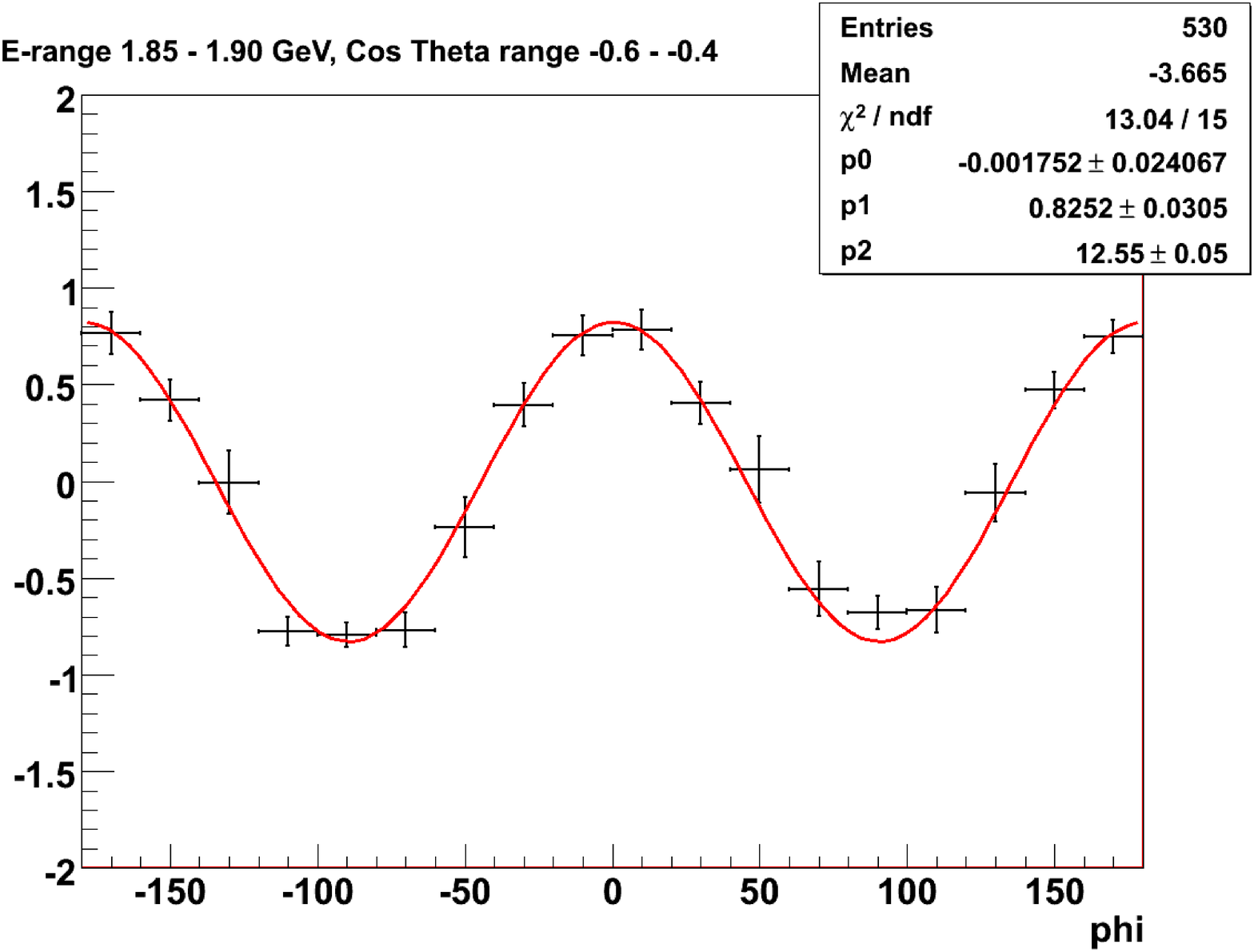}}
\caption{\label{f5} $\frac{N_{\perp}-N_{\parallel}}{N_{\perp}+N_{\parallel}}$  vs. $\phi$ angle as defined in (\ref{simple_asym_eq}), fitted with $\Sigma P \cos2\phi$.}
\end{minipage} \hfill
\begin{minipage}[t]{0.48\textwidth}
\centerline{\epsfysize 4.7 cm
\epsfbox{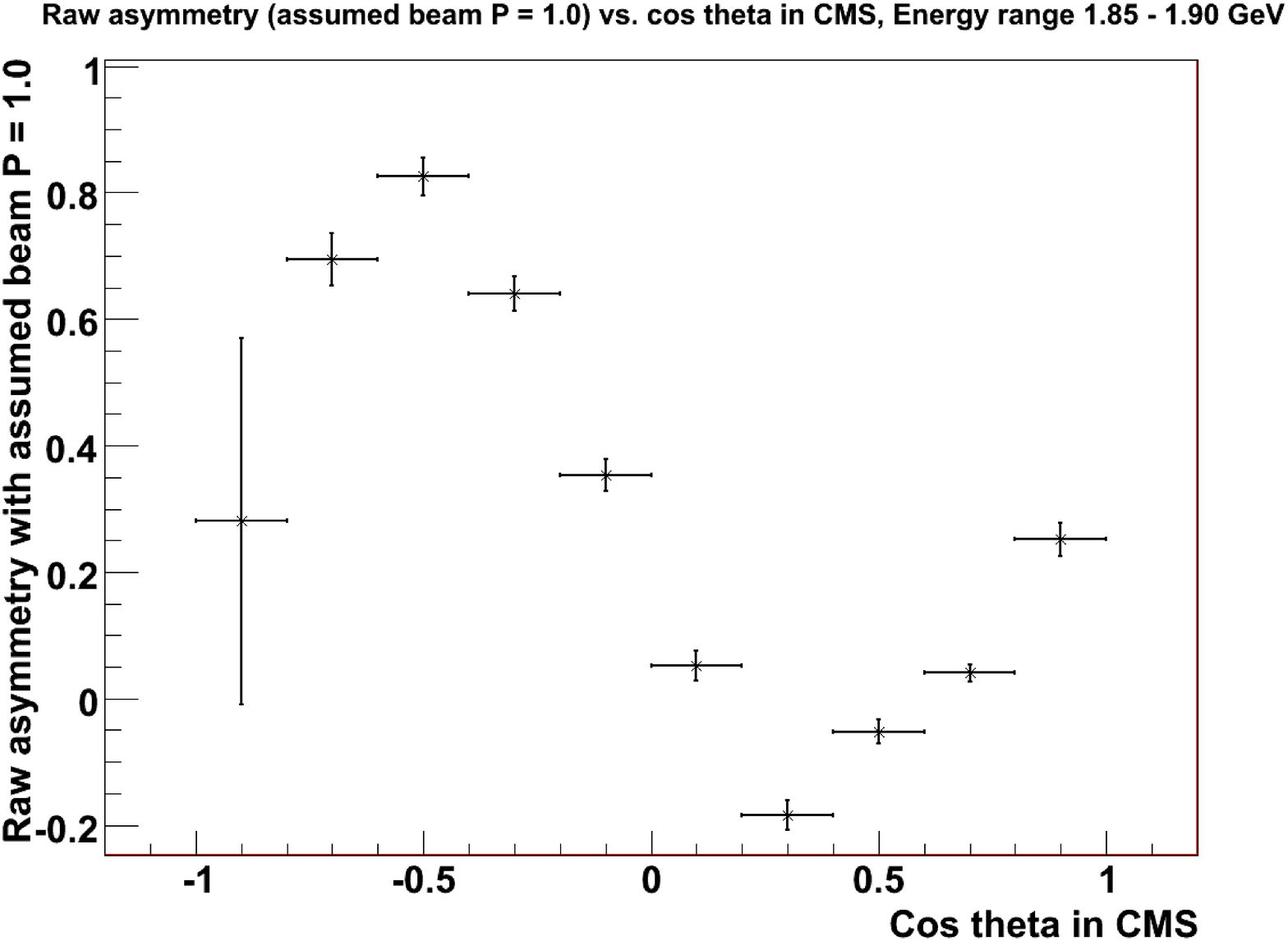}}
\caption{\label{f6} Very preliminary measurement of $\Sigma$, assumed polarisation, {\it P} = 1. Only the statistical error bars are shown.}
\end{minipage}
\end{center}
\end{figure}

As can be seen from (\ref{asym_eq}), the beam asymmetry is extracted from a $\cos2\phi$ fit to the $\phi$-distribution. $\phi$ is, as before, the angle between the reaction plane and the photon polarisation plane. To make the following simplification possible, we need to keep the plane from which we measure $\phi$ constant, so we choose to always measure from the horizontal polarisation plane, which is also the horizontal plane of CLAS in the lab frame.  In order to reduce systematics, the polarisation plane was rotated between two orthogonal directions during the experiment, with $\phi$ being always measured from the horizontal plane of CLAS, simplifying the expression thus (Fig.~\ref{f5}): 
\begin{equation*}
N_{\parallel}={\sigma}_0(1-P \Sigma \cos2\phi)
\end{equation*}
\begin{equation*}
N_{\perp}={\sigma}_0(1+P \Sigma \cos2\phi)
\end{equation*}
\begin{equation}
\Sigma P \cos2\phi=\frac{N_{\perp}-N_{\parallel}}{N_{\perp}+N_{\parallel}}
\label{simple_asym_eq}
\end{equation}
where $N_{\parallel}$ is the differential cross-section from (\ref{asym_eq}) using a parallel beam polarisation and $N_{\perp}$, consequently, a perpendicular polarisation. We make the assumption that $P_{\parallel}=P_{\perp}=P$. Calibrations are still underway, but a very preliminary evaluation of $\sim$ 0.4$\%$ of the data indicates a tiny statistical uncertainty (Fig.~\ref{f6}). It is our goal to reduce systematics, which are currently very large, to $\sim$ 5$\%$.

\section{Prospects}

The determination of the Beam Asymmetry, $\Sigma$, from a linearly polarised photon beam incident on a neutron (in a liquid deuterium target) is underway. Early analysis shows that the data quality is good, statistical uncertainty is tiny and a sizeable asymmetry can be seen at backward angles. A full analysis is to follow soon in scattering angle range 20$^\circ$ - 145$^\circ$ (in CMS) and entire photon energy range 1.1-2.3 GeV. The measurement promises to greatly expand the sparse world data-set on the neutron and aid in constraining the determination of the reaction amplitudes in pion photoproduction thus providing strong constraints on the nucleon excitation spectrum.

\end{document}